\documentclass[journal = apchd5]{achemso}

\usepackage{achemso}
\usepackage[utf8]{inputenc}

\title{Self-probed ptychography from semiconductor high-harmonic generation}

\author{Sven Fröhlich}
\affiliation{Université Paris-Saclay, CEA, CNRS, LIDYL, 91191 Gif-sur-Yvette, France}

\author{Xu Liu}
\affiliation{Université Paris-Saclay, CEA, CNRS, LIDYL, 91191 Gif-sur-Yvette, France}
\alsoaffiliation{Imagine Optic, 18, rue Charles de Gaulle, 91400 ORSAY, France}

\author{Aimrane Hamdou}
\affiliation{Université Paris-Saclay, CEA, CNRS, LIDYL, 91191 Gif-sur-Yvette, France}

\author{Alric Meunier}
\affiliation{Université Paris-Saclay, CEA, CNRS, LIDYL, 91191 Gif-sur-Yvette, France}

\author{Mukhtar Hussain}
\affiliation{Université Paris-Saclay, CEA, CNRS, LIDYL, 91191 Gif-sur-Yvette, France}
\alsoaffiliation{GoLP, Instituto de Plasmas e Fusão Nuclear, Instituto Superior Técnico, Universidade de Lisboa, 1049-001 Lisboa, Portugal}

\author{Mathieu Carole}
\affiliation{Université Paris-Saclay, CEA, CNRS, LIDYL, 91191 Gif-sur-Yvette, France}

\author{Shatha Kaassamani}
\affiliation{Université Paris-Saclay, CEA, CNRS, LIDYL, 91191 Gif-sur-Yvette, France}

\author{Marie Froidevaux}
\affiliation{LOA, ENSTA ParisTech, CNRS, Ecole Polytechnique, UMR 7639, 828 Boulevard des Maréchaux, 91120 Palaiseau, France}

\author{Laure Lavoute}
\affiliation{Novae, 15 Rue Sismondi, 87000 Limoges, France}

\author{Dmitry Gaponov}
\affiliation{Novae, 15 Rue Sismondi, 87000 Limoges, France}

\author{Nicolas Ducros}
\affiliation{Novae, 15 Rue Sismondi, 87000 Limoges, France}

\author{Sébastien Février}
\affiliation{Novae, 15 Rue Sismondi, 87000 Limoges, France}
\alsoaffiliation{Université de Limoges, CNRS, XLIM, UMR 7252, 87000 Limoges, France}

\author{Philippe Zeitoun}
\affiliation{LOA, ENSTA ParisTech, CNRS, Ecole Polytechnique, UMR 7639, 828 Boulevard des Maréchaux, 91120 Palaiseau, France}

\author{Milutin Kovacev}
\affiliation{Leibniz Universität Hannover, Institut für Quantenoptik, Welfengarten 1, D-30167, Hannover, Germany}

\author{Marta Fajardo}
\affiliation{GoLP, Instituto de Plasmas e Fusão Nuclear, Instituto Superior Técnico, Universidade de Lisboa, 1049-001 Lisboa, Portugal}

\author{Willem Boutu}
\affiliation{Université Paris-Saclay, CEA, CNRS, LIDYL, 91191 Gif-sur-Yvette, France}

\author{David Gauthier}
\email{david.gauthier@cea.fr}
\affiliation{Université Paris-Saclay, CEA, CNRS, LIDYL, 91191 Gif-sur-Yvette, France}

\author{Hamed Merdji}
\affiliation{Université Paris-Saclay, CEA, CNRS, LIDYL, 91191 Gif-sur-Yvette, France}
\alsoaffiliation{LOA, ENSTA ParisTech, CNRS, Ecole Polytechnique, UMR 7639, 828 Boulevard des Maréchaux, 91120 Palaiseau, France}

\begin{document}

\begin{abstract}
We demonstrate a method to image an object using a self-probing approach based on semiconductor high-harmonic generation. On one hand, ptychography enables high-resolution imaging from the coherent light diffracted by an object. On the other hand, high-harmonic generation from crystals is emerging as a new source of extreme-ultraviolet ultrafast coherent light. We combine these two techniques by performing ptychography measurements with nano-patterned crystals serving as the object as well as the generation medium of the harmonics. We demonstrate that this strong field in situ approach can provide structural information about the object. With the future developments of crystal high harmonics as a compact short-wavelength light source, our demonstration can be an innovative approach for nanoscale imaging of photonic and electronic devices in research and industry.
\end{abstract}


\section{Introduction}
\begin{figure}
    \centering
    \includegraphics[width=0.75\textwidth]{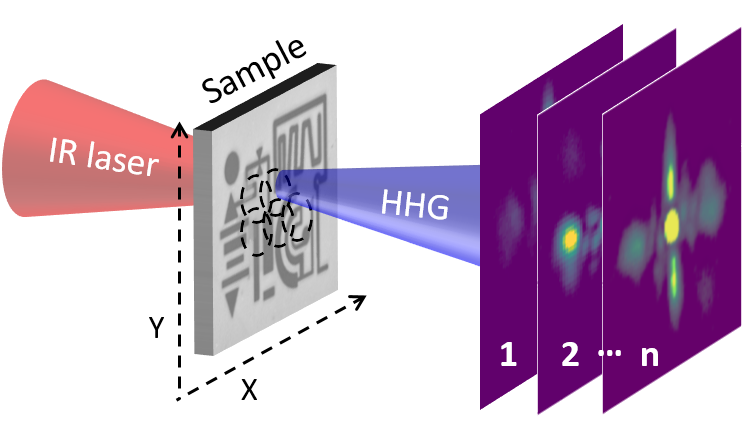}
    \caption{Depiction of the setup used for ptychographic in situ HHG imaging. The incident laser is focused on the rear end of the nanopatterned crystal and generates harmonics directly in the structure. The far field diffraction pattern is then captured by a CCD. After moving the sample the procedure is repeated for the next position until a full array of n diffraction patterns is collected.}
    \label{fig:setup}
\end{figure}

Since its first observation \cite{Ghimire2011} high-harmonic generation (HHG) from band gap solids has been an intensively studied topic that opens up to a wide range of applications. The all-solid-state nature of the generating medium enables more compact UV to extreme-ultraviolet (XUV) HHG sources. Additionally, the harmonics generated from the strong field interaction with the solid are highly sensitive to its electronic band structure, making them an excellent observable for structural properties of the material \cite{Vampa2015}. Furthermore, the visualization of the electron currents associated to the fundamental mechanisms of the strong field process can pave the way for applications in future petahertz optoelectronics \cite{krausz_attosecond_2014,schoetz_perspective_2019,sederberg_attosecond_2020} or light induced topology \cite{gauthier_orbital_2019,silva_topological_2019}. Over the past years high harmonics have been observed in a multitude of materials ranging from 3D bulk \cite{Ghimire2011,luu_extreme_2015,kim_generation_2017,nefedova_enhanced_2021} to 2D crystals \cite{yoshikawa_high-harmonic_2017,liu_high-harmonic_2017,hafez_extremely_2018}. Out of the many possibilities, nanostructuring of the material is especially interesting as it allows to increase the generated harmonic yield and directly control the spatial structure of the generated harmonics  \cite{sivis_tailored_2017,Franz2019,gauthier_orbital_2019,Liu2020}. The sub-wavelength sensitivity of the generating beam to those nanostructures can be further exploited to monitor nanoscale spatial amplitude and phase modulations of the crystal itself. \newline
Coherent diffractive imaging (CDI) is a powerful technique that has been widely used for nanoscale imaging since its first experimental demonstration \cite{miao_extending_1999}. By solving the so-called phase problem, using a phase retrieval algorithm, this lensless technique can reach a spatial resolution limited only by the wavelength. Ptychography was developed to generalize CDI to extended objects \cite{pfeiffer_x-ray_2018}. It uses multiple coherent diffraction patterns originating from overlapping light probes on the structure. The image reconstruction of the whole structure is performed using the redundant information provided by these multiple diffraction patterns using an iterative algorithm \cite{rodenburg_phase_2004,thibault_high-resolution_2008}. Stunning demonstrations have been performed using synchrotron radiation or HHG-gas sources in biology and solids state physics \cite{pfeiffer_x-ray_2018,holler_high-resolution_2017, giewekemeyer_quantitative_2010,shi_soft_2016,baksh_wide-field_2016,loetgering_tailoring_2021}. \newline
In this work we demonstrate the use of HHG from solids as an integrated source of light to image a sample using ptychography. The key here is to combine in a 2-in-1 configuration the generation medium and the object, as illustraded in Figure \ref{fig:setup}. In this configuration, the ptychographic HHG probe is intricated with the object. Indeed HHG from solids is a spatially coherent light source that qualifies also well for coherent diffractive imaging as was demonstrated recently \cite{Franz2019}. The conservation of the coherence properties during the generation process makes the harmonics a suitable intrinsic self-probe for ptychography. Compared to a previous work that uses the perturbative second harmonic generation \cite{Odstrcil}, the non-perturbative high-harmonic generation allows to image the object with significant structural information thanks to the sensitivity of the strong field emission process to the material properties. Finally, another interest of the ptychographic approach is the compatibility with tight focusing. This enables low average power femtosecond fiber lasers to reach sufficient intensities for HHG in solids and opens up possibilities for extremely compact and affordable arrangements.

\section{Methods}
In our experiments we use a commercial all-fiber laser delivering 90\,fs pulses with 10\,nJ at a central wavelength of 2100\,nm and 18.66 MHz repetition rate, having linearly polarized purely single mode output beam \cite{NOVAE}. The laser beam is focused at the rear side of the sample by an aspheric lens with 3\,cm focal length, reaching a focal spot size of 6.5\,µm (FWHM), with intensities of about 0.2\,TW/cm\(^{2}\). Self-focusing during the propagation of the beam in the sample can additionally decrease the laser focus and furhter increase the peak intensity. The fifth  harmonic order at a wavelength of 420 nm is selected using a bandpass filter set just before the CCD camera. Because of mechanical limitations and of the limited chip size of the CCD, a microscope objective (numerical aperture (NA) of 0.75) is set after the sample collecting the wide angle diffraction patterns on the detector. It images the far field diffraction pattern from a virtual plane located about 200\,µm after the sample.
The raw ptychographic data are pre-processed before the reconstruction, this includes: i) binning and cropping of the diffraction pattern, ii) subtraction of background noise and iii) correction for the projection of the Ewald sphere onto the flat detector. As a consequence of this pre-processings, the effective NA of the diffraction data set is limited to 0.5. The calibration of the microscope objective magnification (of about 20) is performed by comparing the experimental data to numerical simulations. The ptychographic data were then used to perform the amplitude and phase image reconstruction of the nano-patterned samples using the PyNX library \cite{Favre-Nicolin:2020}.
\newline

\section{Results and discussion}
\subsection{Nanopatterned silicon sample}
\begin{figure}
    \centering
    \includegraphics[width=0.45\textwidth]{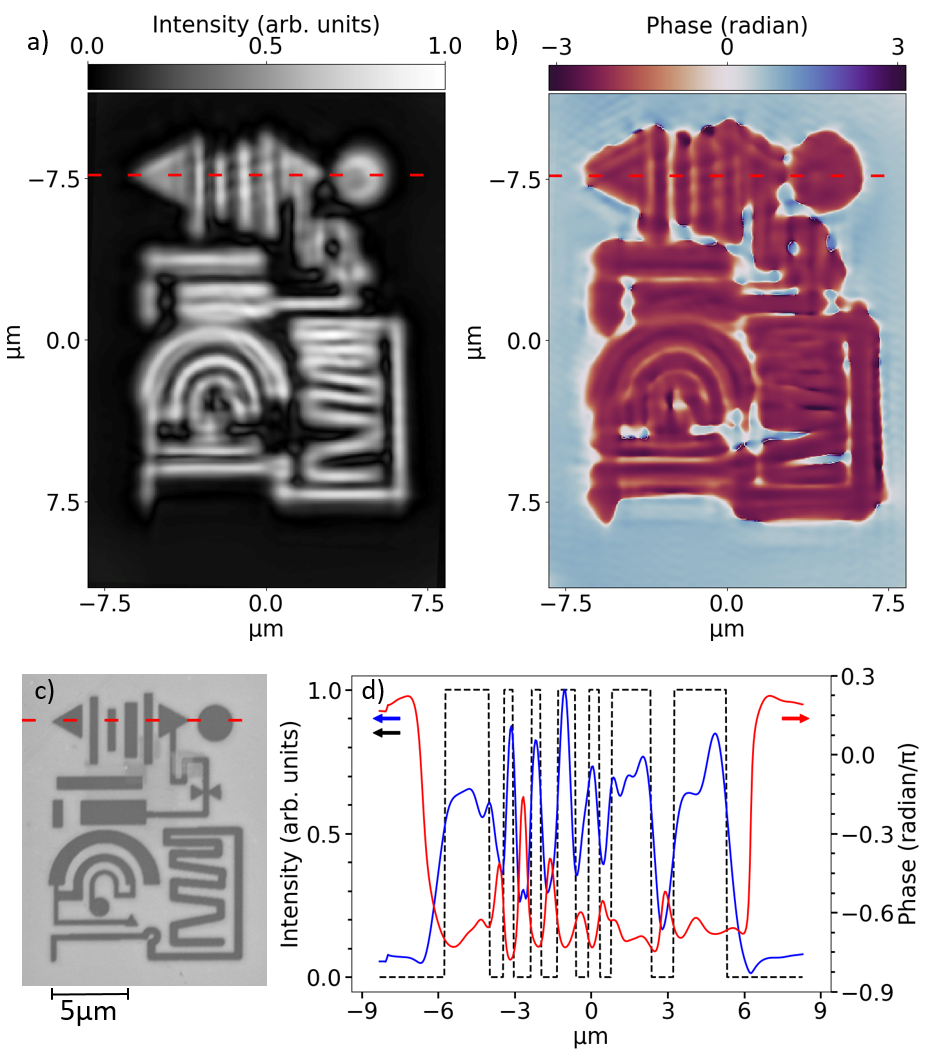}    
    \caption{Reconstruction of the silicon sample. The amplitude (a) and phase (b) of 10 averaged independent reconstructions of the field at the exit of the structure. A scanning electron microscopy image of the patterned structure is shown in (c). (d) shows a profile along the dashed lines in (a) and (b). The black dashed lines in (d) represent the expected positions of the structure edges.}
    \label{fig:Ptycho4}
\end{figure}

\begin{figure}
    \centering
    \includegraphics[width=0.8\textwidth]{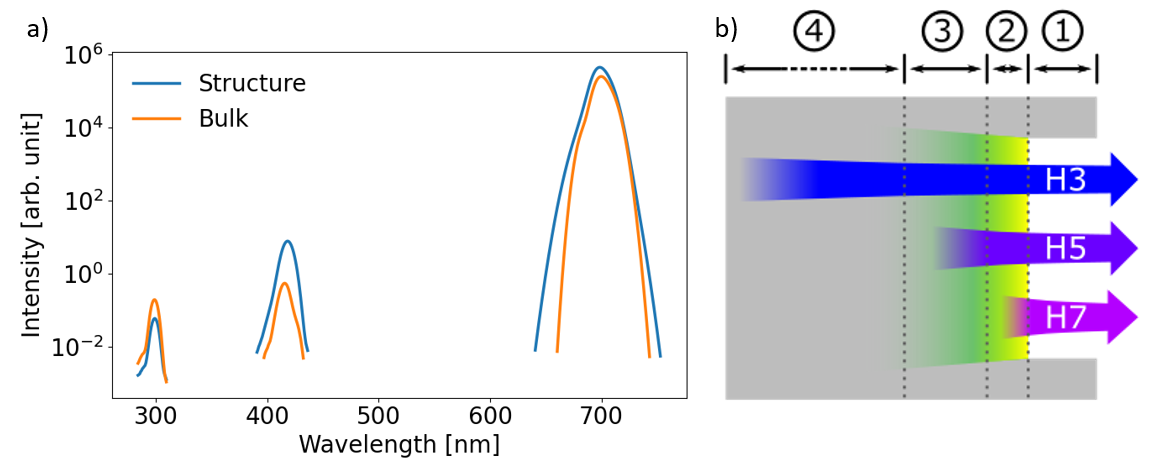}
    \caption{(a) Spectra recorded at a position inside the patterned structure (blue) and an unstructured area of bulk silicon (orange). 
    (b) Illustration of the depth of the source of emission for the first three harmonics that effectively propagate outside the crystal. In the structured part, first comes an area where material has been removed with the FIB (1). Next is an area where the crystal structure is transferred into an amorphous phase (2) followed by an area where the Ga implementation leads to a high concentration of crystal defects (3). At large depth only the pristine silicon remains (4).}
    \label{fig:Spectra}
\end{figure}

The first structure is patterned onto a 300\,µm (100)-cut silicon sample using a Focused Ion Beam (FIB). The test geometry is a custom mix of features commonly used in scientific and industrial applications (see Figure \ref{fig:Ptycho4} (c)). Its overall size is 18x12\,µm\(^{2}\). The depth is chosen to be about 50\,nm to achieve a well contrasted phase object. The structure is probed in a spiral scanning pattern using 2000 points with an average step size of 500\,nm and with 3 seconds of integration time each. The polarization of the laser is aligned along the short axis of the structure that matches the [110]-axis (\(\Gamma\)K) of the silicon sample. The reconstruction uses 40 iterations of the \textit{alternating projections} algorithm followed by 40 iterations of \textit{maximum likelihood} algorithm retrieving the probe simultaneously with the image of the object. The reconstructions are performed on a GPU using PyNX CUDA supported modules. To avoid artifacts of the reconstruction process, 10 independent reconstructions are averaged. The retrieved probe, that reconstructs the harmonic source, has a size of 2.5\,µm (FWHM). The size reduction in respect to the laser focus results from the self-focusing effect and from the harmonic nonlinear conversion. \newline
An example of a reconstruction in amplitude and phase of the sample is shown in Figure \ref{fig:Ptycho4}. It retrieves all the  features of the sample. A resolution of 800\,nm is estimated using a 10\%-90\% edge criterion on the rightmost edge of the amplitude in Figure \ref{fig:Ptycho4} (d), which is less than twice the theoretical resolution limit for this setup ($\textrm{d} = \frac{\lambda}{2 \textrm{NA}} = 420\textrm{\,nm}$). The comparison of the cut through the reconstructed structure with a cut through the original pattern shows a good overlap. We note the presence of unexpected amplitude modulations in the reconstruction. This effect takes its origin in the propagation of H5 in the structure \cite{zayko_coherent_2015} and is qualitatively confirmed by \textit{finite difference time domain} (FDTD) simulations. Moreover, the simulation of the propagation of the fundamental laser confirms no significant confinement in the 50\,nm depth nanostructures. The overall phase profile shows a contrast of 0.95\(\pi\), close to the designed phase difference of \(\pi\). However, The amplitude reveals an increase of the coherent signal from the patterned area compared to the unpatterned parts of the sample by tenfold that does not match the expected design of a phase object. Comparative spectroscopic measurements of the HHG emission between the structured sample and the bulk material have been conducted and have confirmed this increase of the harmonic 5 signal from the patterned sample (Figure \ref{fig:Spectra} (a)). This behaviour is also visible for H3, which is slightly increased. In contrast H7, the first above band gap harmonic, is decreased significantly. \newline
This increase of below band gap harmonics was observed by \textit{Sivis et al} \cite{sivis_tailored_2017} for FIB induced Ga\(^{+}\) ion implantation as well. However the ion dose (at 30 keV) used to etch the structures in our work is larger, leading to a higher relative fraction of Ga ions of up to 0.1 \cite{gnaser_focused_2008}. The process also generates secondary effects like the amorphization of the crystal structure close to the surface \cite{gnaser_focused_2008,lehrer_limitations_2001,schrauwen_iodine_2007}. This effect can create tails of localized states into the bands \cite{horowitz_validity_2015}, while the doping, as well as the crystal defects (e.g. vacancies) occurring alongside, add defect bands in the band gap or can even modify the band gap itself. The increase in signal for the below band gap harmonics (H3 and H5) could result from an increased probability of interband transitions (photon emission from the electron-hole recombination) from those additional states present in the band gap of the pristine silicon. This is particularly true for H5 whose energy is close to the band gap energy. Theoretical studies have reported such enhancement for below band gap harmonics, in the case of pure doping and vacancies \cite{pattanayak_influence_2020,mrudul_high-harmonic_2020}. However, crystal amorphization has been studied theoretically and experimentally and have been found to significantly lower the HHG yield \cite{orlando_high-order_2018, you_high-harmonic_2017}. Thus, the enhancement of below band gap harmonics is counter-acted by the overall HHG signal reduction due to the amorphization. Additionally, each harmonic experiences a different absorption by the material. This absorption also depends on the changes induced by the FIB and consequently changes with depth. All together, the measured part for each harmonic is generated up to a different depth and experiences a different sensitivity to the material changes as sketched Figure \ref{fig:Spectra} (b). For our case H3 is generated at a large depth and is therefore less affected than H5 and H7 that are generated closer to the surface.

\subsection{Spiral zone plate patterned on ZnO}
\begin{figure}[ht]
    \centering
    \includegraphics[width=0.7\textwidth]{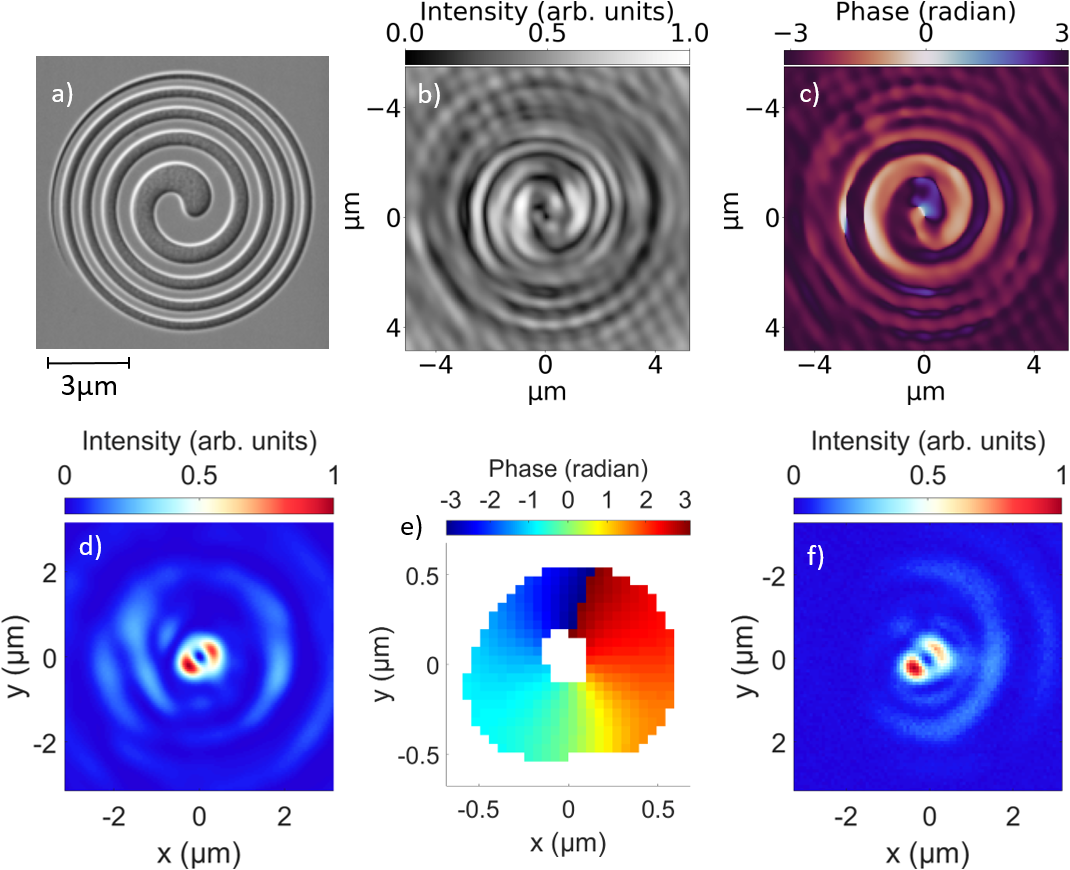}
    \caption{(a) Scanning electron microscopy image of the spiral zone plate. Reconstructions of the intensity (b) and phase (c) distribution of H5 after the SZP. Intensity (d) at the focusing plane of the SZP, i.e. from the numerical propagation over 6 µm after the sample, reconstructing the OAM beam. (e) The corresponding phase for the central ring shape. (f) A prior direct measurement at the SZP focal plane published in [\cite{gauthier_orbital_2019}].}
    \label{fig:SZP}
\end{figure}
As a second test of the capabilities of ptychography with HHG from solids, we studied a 10x10\,µm\(^{2}\) spiral zone plate (SZP) patterned on a 500\,µm thick zinc oxide (ZnO) crystal. The structure is designed to generate a Laguerre-Gaussian beam carrying an orbital angular momentum (OAM) \(\ell = 1\) without the need to pre-shape the driving IR field \cite{gauthier_orbital_2019}. An example of an SZP is shown in Figure \ref{fig:SZP} (a). It is designed to be a pure phase object combining the phase profiles of helical and spherical waves to induce the OAM and focus the beam 6\,µm after the structure. The harmonic used for the measurements is again H5 which lies below the band gap energy of ZnO (3.3\,eV) and is hence generated in the bulk before being diffracted by the SZP. The etching depth is chosen to be 210\,nm resulting in a phase step of \(\pi\). Measurements are performed under similar conditions as for the silicon sample, but scanning the sample in a grid pattern with an average step size of 400\,nm. We attempt to retrieve the induced OAM from the patterned structure based on the ptychography measurements similar to the approach presented by \textit{Vila Comamala et al} \cite{Vila-Comamala}. The data were reconstructed using a combination of algorithms with a total of 210 iterations of \textit{alternating projections} and 10 iterations of \textit{difference map}, performed using the CPU supported module of PyNX. \newline 
The results of the complex field at the exit of the SZP in amplitude and phase are shown in Figure \ref{fig:SZP} (b) and (c). The expected well contrasted binary phase profile is retrieved within the boundaries of the scanned area. This object behaves closely as a pure phase object with a low amplitude contrast between the patterned and non-patterned areas, a behaviour that can be explained by the higher resistance of ZnO to material alterations in the FIB process \cite{kucheyev_ion-beam-produced_2003}. Amplitude vs. phase contrast are clearly target dependant, emphasising the potential of this technique for elemental mapping. However, as the ZnO sample features a much deeper structure than the silicon, the amplitude picture shows a double spiral as a results of the strong propagation effects expected for both the fundamental and H5, which is in qualitative agreement with FDTD simulations. The stronger confinement effect in the structure from the vertical polarization is particularly visible. The numerical propagation of the reconstructed complex field to the focal plane of the SZP is shown in Figure \ref{fig:SZP} (d) and (e). It matches the result from our previous experiments in [\cite{gauthier_orbital_2019}] (reproduced in Figure \ref{fig:SZP} (f)). We retrieve the same aberration due to the limited number of illuminated diffracting zones from the structure. In addition, we access the phase profile that confirms the OAM \(\ell = 1\) carried by the focused beam.

\section{Conclusion}
In conclusion, we demonstrated that self-probed ptychography allows for robust reconstruction of various structures and can be a valuable asset for the characterization of nanostructures or meta-optics in the context of HHG from solids. This approach can be scaled towards the cutoff of HHG from solids, reaching currently 50\,nm \cite{luu_extreme_2015}, which translates directly into a gain of resolution. Additionally, the sensitivity of the HHG mechanisms based on the non-perturbative nonlinear properties of the material to the driving laser field can give access to valuable spectroscopic information from the reconstructed image. In combination with the recent potential in multi-wavelength imaging \cite{multiwavelength_ptycho}, self-probed ptychography could allow a full access to the spatio-spectral image of the sample. Finally, high spatial and temporal resolution could be obtained in conjunction with broadband CDI algorithms \cite{broadband_CDI}.

\section{Funding} We acknowledge the financial support from the PETACom FET Open H2020 grant No. 829153/ OPTOLogic grant No. 899794/ TSAR grant No. 964931 and EIC Open grant No. 101047223-NanoXCAN. We acknowledge the financial support from the French ANR through the grants PACHA (ANR-17-CE30-0008-01), FLEX-UV (ANR-20-CE42-0013-02) and ATTOCOM (ANR-21-CE30-0036-02). We acknowledge the financial support from the French ASTRE program through the “NanoLight” grant. This work was partially supported by the Fundação para a Ciência e Tecnologia (grant No. PD/BD/135224/2017).

\section{Acknowledgments}
We thank Vincent Favre-Nicolin and Kadda Medjoubi for the valuable advice and discussions regarding the reconstructions and PyNX library. We thank Juliana Tiburcio, Reda Berrada and Vijay Sunuganty for the fruitful discussions. We acknowledge Franck Fortuna and the use of the SEM-FIB facility of the Institut des Sciences Moléculaires d'Orsay, for the sample structurations.

\section{Notes} The authors declare no competing financial interest.

\section{Data availability} The data are available from the corresponding author upon reasonable request.

\bibliography{sample}

\end{document}